\documentclass[12pt,twocolumn]{article}
\usepackage[T1]{fontenc}
\usepackage{amsmath, amssymb, amsfonts}
\usepackage{graphicx}
\usepackage{hyperref}
\usepackage{listings}
\usepackage{natbib} 
\usepackage{color}
\usepackage{authblk}
\usepackage{fancyhdr}
\usepackage{times}

\definecolor{codegreen}{rgb}{0,0.6,0}
\definecolor{codegray}{rgb}{0.5,0.5,0.5}
\definecolor{codepurple}{rgb}{0.58,0,0.82}
\definecolor{backcolour}{rgb}{0.95,0.95,0.92}

\lstdefinestyle{mystyle}{
  backgroundcolor=\color{backcolour}, 
  commentstyle=\color{codegreen},
  keywordstyle=\color{magenta},
  numberstyle=\tiny\color{codegray},
  stringstyle=\color{codepurple},
  basicstyle=\ttfamily\footnotesize,
  breakatwhitespace=false,         
  breaklines=true,                 
  captionpos=b,                    
  keepspaces=true,                 
  numbers=left,                    
  numbersep=5pt,                  
  showspaces=false,                
  showstringspaces=false,
  showtabs=false,                  
  tabsize=2
}

\lstdefinelanguage{JavaScript}{
    keywords={break, case, catch, continue, debugger, default, delete, do, else, finally, for, function, if, in, instanceof, new, return, switch, this, throw, try, typeof, var, void, while, with},
    sensitive=true,
    comment=[l]{//},
    morecomment=[s]{/*}{*/},
    morestring=[b]\',
    morestring=[b]",
    ndkeywords={class, export, boolean, throw, implements, import, this},
    keywordstyle=\color{blue}\bfseries,
    ndkeywordstyle=\color{cyan}\bfseries,
    identifierstyle=\color{black},
    commentstyle=\color{green}\ttfamily,
    stringstyle=\color{red}\ttfamily,
    moredelim=[il][\textcolor{orange}]{\$},
    moredelim=[is][\textcolor{orange}]{\%\%}{\%\%}
}

\title{\textbf{Large Language Models for Behavioral Economics \\[2pt] Internal Validity and Elicitation of Mental Models}}

\author{Brian Jabarian\footnote{Contact: \href{mailto:brian.jabarian@chicagobooth.edu}{brian.jabarian@chicagobooth.edu}. This article is an entry under preparation for the Elgar Encyclopedia of Experimental Social Science edited by Bereket Kebede. I thank the editor Bereket Kebede for helpful suggestions. I am grateful to Gary Charness, Rishane Dassanayake and John List, for helpful comments. I thank Rishane Dassanayake for excellent research assistance. I thank Effective Ventures US for financial support.}\\
\vspace{-8pt}
University of Chicago Booth Business School \\
\vspace{5pt}
\textit{\small Entry under preparation for the Elgar Encyclopedia of Experimental Social Science (Forthcoming)}}
\date{June 28, 2024}

\begin{document}

\maketitle
\pagestyle{fancy}
\fancyhf{}
\fancyhead[RO,LE]{\textit{Forthcoming in the Elgar Encyclopedia of Experimental Social Science}}

\begin{abstract}
    In this article, we explore the transformative potential of integrating generative AI, particularly Large Language Models (LLMs), into behavioral and experimental economics to enhance internal validity. By leveraging AI tools, researchers can improve adherence to key exclusion restrictions and in particular ensure the internal validity measures of mental models, which often require human intervention in the incentive mechanism. We present a case study demonstrating how LLMs can enhance experimental design, participant engagement, and the validity of measuring mental models. 
\end{abstract}

\section{Introduction}

The integration of generative artificial intelligence (AI), in particular, Large Language Models (LLMs), into behavioral and experimental economics can represent a transformative scientific advancement. In \cite{Charness_Jabarian_List_2023}, we explore how generative AI can augment the process of social scientific discovery for each stage of the scientific process (design, implementation, and analysis). In this article, we focus on how researchers can leverage LLMs to optimize internal validity. First, we discuss how researchers can use specific AI tools to optimize adherence to exclusion restrictions. Second, we discuss an early case of how AI and LLMs were leveraged to optimize such adherence and facilitate the incentivized study of new behavioral variables that are usually hard to track: thinking styles in a natural generation of digital environments for lab-in-field experiments. We show how these tools can be leveraged to create engaging storytelling environments, ensure incentive compatibility, and monitor undesirable participant behavior in online experiments. 

\section{Leveraging LLMs to Optimize Exclusion Restrictions}

In social science research, maintaining the internal validity of experiments and adherence to the four exclusion restrictions – observability, compliance, stable unit treatment value assumption (SUTVA), and statistical independence – is essential \citep{List_2023} to recover causal effects in experiments. This section delves into how AI can be leveraged to enhance adherence to these restrictions, ensuring more accurate and reliable social science research. 

First, observability requires that the treatment and control conditions are clearly distinguishable and consistently applied. This means ensuring that all participants remain in the study and that their outcomes, treatment assignments, and characteristics are observable and accurately recorded. AI can enhance observability by generating clear, concise instructions and comprehension checks tailored to the participant's language proficiency and education level. This can be done by generating multiple versions of instructions and comprehension checks and using LLMs like GPT-4 to assist in critiquing and refining them at scale \citep{Saunders_et_al_2022}. This customization ensures that participants understand the experimental setup and the differences between treatment and control conditions. In particular, state-of-the-art models like GPT-4, which are trained on billions of data points, can customize the language, tone, and complexity of instructions to suit participants with different levels of language proficiency \citep{Charness_Jabarian_List_2023}. AI can also monitor participant engagement in real-time by, for example, offering intermittent check-ins via a chatbot interface, ensuring that they are focused on the task and not distracted. This helps maintain the integrity of the experimental conditions and ensures that the treatments are consistently applied.

Second, compliance is critical in ensuring that participants adhere to the experimental protocol. Non-compliance can lead to significant measurement errors, jeopardizing the validity of causal inferences, especially in online experiments \citep{Gillen_Snowberg_Yariv_2019}. This aligns with the Complete Compliance exclusion restriction, which assumes that every unit assigned to a treatment or control group adheres to that assignment. AI, particularly LLMs, can be employed to enhance compliance in several ways. Firstly, AI chat assistants can provide real-time support to participants during the experiment, answering questions and clarifying instructions. This helps ensure that participants understand the tasks and adhere to the experimental protocols, reducing instances of non-compliance. For example, chatbase.co\footnote{See \url{https://www.chatbase.co/}} provides a service that allows researchers to connect data sources, train AI chatbots, and deploy them by embedding a simple HTML code in two clicks. Due to its scalability and flexibility, incorporating this approach could become a common practice for future online experiments, leading to significant advancements and innovations in online research and surveys \citep{Charness_Jabarian_List_2023}. Furthermore, AI can monitor participant behavior in real time, detecting deviations from the protocol. For instance, LLMs, in conjunction with other software engineering methods, can track whether participants open new tabs, switch windows, or plagiarise written content, behaviors that might indicate non-compliance \citep{Jabarian_Sartori_2023}. If detected, AI can prompt participants to return to the task or amend their responses, ensuring adherence to the experimental conditions.

Third, SUTVA posits that the treatment given to one participant should not affect the outcomes of another. Ensuring SUTVA is challenging, particularly in online experiments exploring concepts in game theory, auction theory, or social norms where participants might interact or influence each other indirectly. In these cases, AI can help maintain SUTVA by creating highly controlled virtual environments where participants interact with simulated agents instead of real people. For example, EVE (Experiments in Virtual Environments)\footnote{See \url{https://cog-ethz.github.io/EVE/}} is an open-source framework that allows researchers to create virtual 3D environments to run experiments \citep{Grubel_Weibel_Jiang_Holscher_Hackman_Schinazi_2017}. For the kinds of experiments mentioned above, integrating AI into these environments may help ensure the SUTVA condition is met by reducing the risk of cross-participant influence and helping to maintain the integrity of individual treatments while still allowing researchers to glean behavioral insights. AI could also manage and automate randomization processes, ensuring that treatment assignments are genuinely random and not influenced by external factors. Re-randomization techniques further ensure that any potential biases in treatment assignments are mitigated.

Fourth, statistical independence requires that the assignment of treatments is independent of any other factors that could influence the outcome. It implies that the treatment assignment is orthogonal to the potential outcomes, eliminating selection bias and ensuring that the observed difference in outcomes between treatment and control groups can be attributed solely to the treatment effect. AI could, further down, ensure independence by automating the randomization process, ensuring that external factors do not influence treatment assignments. Significantly, AI could facilitate the implementation of transparent algorithms used to generate random assignments, which could be audited to ensure fairness and independence. This guarantees that each participant has an equal probability of being assigned to any treatment condition, maintaining the integrity of the randomization process. AI could also analyze the data for patterns that might indicate bias in the treatment assignments. If detected, the AI could adjust the randomization process or flag the data for further review. In addition, for experiments requiring matched pairs or groups, AI can ensure that matching is done accurately and without introducing biases. Machine learning/AI models are uniquely suited for this task due to their ability to handle high-dimensional data while avoiding multicollinearity. This allows AI-powered matching algorithms to group participants by factors that are less directly observable to experimenters and may be highly correlated. By thoroughly analyzing participant data, AI can create matched pairs or groups that are equivalent in all relevant respects, thereby upholding the statistical independence condition.

\section{Leveraging LLMs for Eliciting Mental Models}

LLMs can be leveraged to explore mental models and behaviors in complex experimental designs, such as in \cite{Jabarian_Sartori_2023} or \cite{Arrieta_Nielsen_2024}. In this latter, we explore whether exposure to various storytelling types can act as mental nudges, prompting agents who are faced with novel, complex societal issues (issues where no objective truth is present to inform opinions, e.g., abortion, immigration, civil rights) to transition from a ‘naive thinking’ to a ‘critical thinking’ reasoning style.  The experiment involved 860 participants from the US, screened down to 725 after ensuring attention and quality. Participants were randomly assigned to different storytelling types mimicking social media and traditional media platforms (e.g., Twitter, Facebook, Newspaper). A significant part of the experimental design of this study leverages LLMs and Javascript to ensure the internal validity of the experiment. 

\begin{figure}[h!]
    \centering  
    \includegraphics[width=\textwidth]{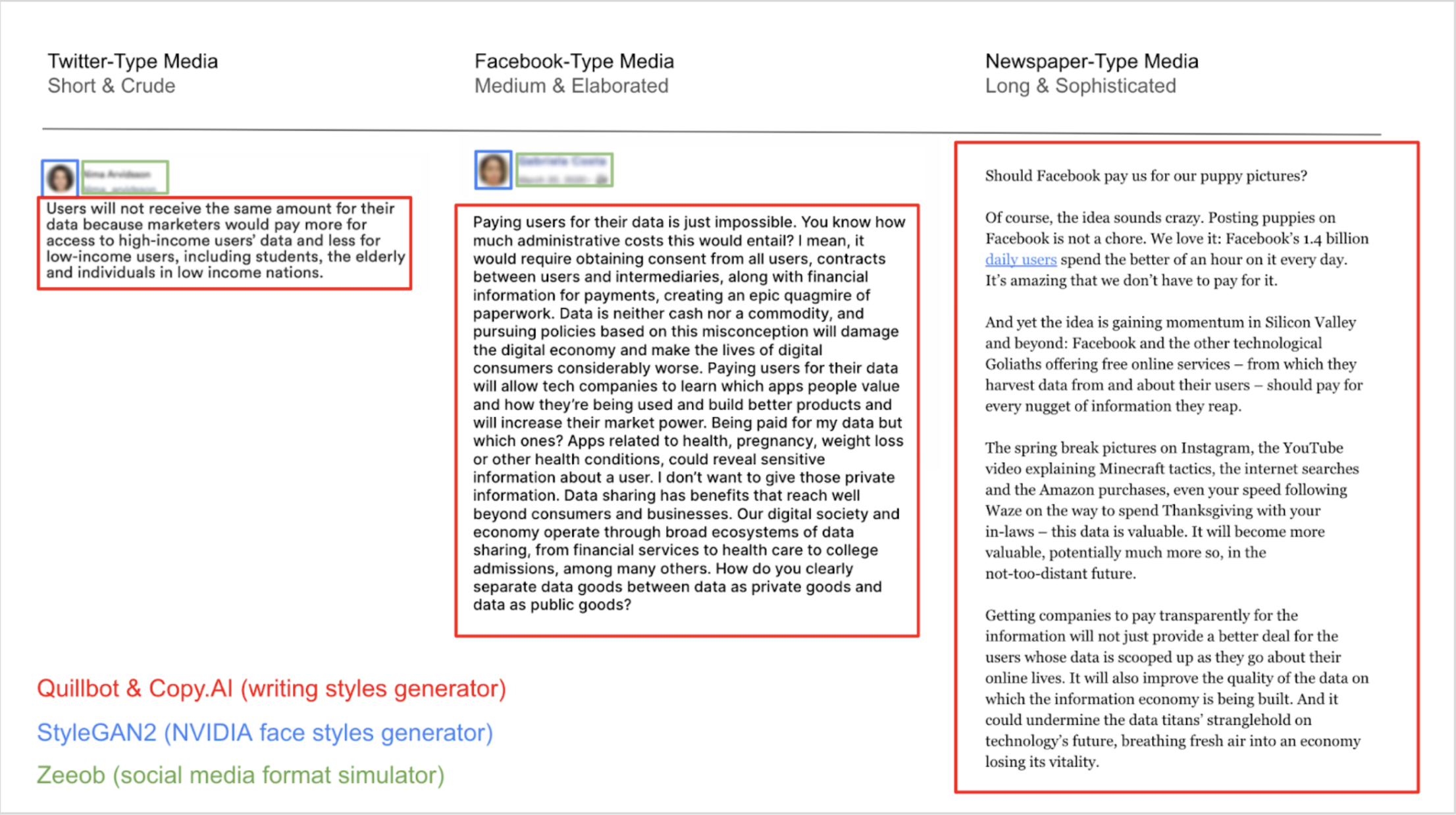}
    \caption{An example of generated user interface using AI tools.}
    \label{fig:1}
\end{figure}

First, LLMs were employed to create engaging storytelling environments. Although no deception was involved, this ensured that the environment felt familiar and natural to participants to optimize engagement. LLM-powered apps such as Quillbot\footnote{See \url{https://quillbot.com/}} and Copy.AI\footnote{See \url{https://www.copy.ai/}} aided in generating the different writing lengths and styles associated with each storytelling type, and StyleGAN2\footnote{See \url{https://github.com/NVlabs/stylegan2}}, an image generation model, was used to generate fake user profile pictures. Zeeob\footnote{See \url{https://zeoob.com/}}, a social media format simulator, was used to generate the UX design elements that are specific to various media formats, such as the fonts, line spacing, color, and other elements. See \autoref{fig:1} for an example of a generated user interface. 

Second, the measurement of critical thinking was incentivized through a transparent and standardized grading system. This system, powered by Grammarly, utilized LLM-generated metrics and reports benchmarked against a large US population. While LLMs facilitated the initial grading, cognitive psychologists were involved in the final evaluation of the critical thinking essays. Each essay was independently assessed by three experts, ensuring a majority grading system that minimized individual biases. This combination of AI and human expertise provided a robust framework for evaluating critical thinking. In particular, the inclusion of LLMs ensured that the grading was consistent, independent, and credible. This is because, since the essays focused on political topics, participants may have provided answers that pandered to the projected political views of psychologist graders. This would have distorted the results. Instead, since compensation was tied to the grade they received from the AI system, which presumably has less expected political bias, we avoid this problem.

Third, JavaScript algorithms were implemented at the different stages of the collection to enhance the data quality. Before the collection, we forced participants to take the experiment on a monitor screen (hence avoiding participants using different devices, which could become a new source of measurement error). Besides, they were required to have Google Chrome installed (and could be installed directly with a provided link) to also minimize the risks that using different browsers to take part in our experiment would become another source of measurement error (since each participant may have their personalized configurations on their browser and each browser rendering the UX design of a given experiment differently). During the experiment, in the incentivized tasks from which participants’ bonuses were determined, we monitored the quality of their inputs. We designed algorithms to check whether participants opened new tabs to access external information that may aid in completing the assigned tasks. They also detect whether participants copied and pasted information by comparing the keystroke count and character count of their responses (see \autoref{fig:2}). After the collection, an LLM-powered detection tool was used to check for plagiarism, ensuring the authenticity of the participants' submissions. At the time of our data collection in 2020, the risk of an LLM-powered detection tool not detecting whether a human or AI wrote it was low, and hence, such tools were efficient in checking for plagiarism. Now, with the difficulties in detecting the authenticity of content generation \citep{Elkhatat_Elsaid_Almeer_2023, Weber-Wulff_Anohina-Naumeca_Bjelobaba_et_al_2023}, live-detection methods (such as monitoring typing speed, vocabulary heterogeneity, and other detectable signals) or like those implemented here, become vital to detect behavioral authenticity and compliance with the instructions (checking live for copy-paste and plagiarism). 

\begin{lstlisting}[language=JavaScript, caption={JavaScript code for Qualtrics survey engine}, label={fig:2}]
Qualtrics.SurveyEngine.addOnload(function()
{
    var body = this.getQuestionContainer();
    var keycount = 0;
    var currentLength = 0;
    var textString = 0;

    var textbox = document.getElementsByClassName("InputText")[0];
    console.log(textbox)
    textbox.style.width = "${e://Field/essayBoxWidth}" + "px";
    textbox.style.height = "${e://Field/essayBoxHeight}" + "px";

    body.addEventListener('keyup', function (e) {
        var key = e.which || e.keyCode;
        keycount++; 
        console.log("Pressed " + key + " | count = " + keycount);
        Qualtrics.SurveyEngine.setEmbeddedData("keycount", keycount);
        textString = textbox.value;
        console.log(textString);
        currentLength = textString.length;
        console.log(currentLength);
        document.getElementById("ccount").innerHTML = currentLength + "/800";
    });
});
\end{lstlisting}

\section{Conclusion}

The application of AI in experimental economics offers unparalleled opportunities to improve the rigor and scope of social science research. By ensuring compliance with key exclusion restrictions, AI tools such as LLMs enhance the internal validity of experiments, providing more accurate and reliable data. The ability to simulate human behavior and generate synthetic behavioral data with AI agents represents a significant methodological advancement, allowing researchers to explore decision-making processes in unprecedented depth. As AI technology continues to evolve, its integration into experimental economics promises to foster innovative research methodologies, yielding deeper insights into human behavior and decision-making. This convergence of AI and experimental economics is poised to revolutionize the field, promoting greater scientific rigor, transparency, and reproducibility.

\bibliographystyle{apalike}
\bibliography{bibliography}

\newpage
\appendix

\section{Appendix}

\end{document}